
\input harvmac.tex

\nref\rBELL{J.S. Bell, Physics {\bf 1} 195 (1964).}
\nref\rEPR{ A. Einstein, B. Podolsky, N. Rosen, Phys. Rev. {\bf 47},777
(1935).}
\nref\rGRIF{   R.B. Griffiths - Amer. J. Phys. {\bf 55},11 (1987).}
\nref\rBOHM{D. Bohm, {\it Quantum Theory}. Prentice-Hall, Englewood Cliffs,
1951.}
\nref\rCLAU{S.J.Freedman and J.F.Clauser, Phys. Rev. Lett. {\bf 28} 938
(1972).}
\nref\rBARG{ V. Bargmann, Ann. Math. {\bf 59} (2), 1 (1954).}
\nref\rBELIN{ J. Belinfante, J. Math. Phys. {\bf 17} No. 3, 285 (1976).}
\nref\rFIVone{ D.I. Fivel, Phys. Rev. Lett. {\bf 67}, 285 (1991).}
\nref\rFIVtwo{ D.Fivel, U.of Md. Preprint UMD-PP-94-133 }
\nref\rFIVthree{ D.Fivel, U.of Md. Preprint UMD-PP-94-131 }
\nref\rBEN{ C.H. Bennett and S.J. Wiesner, Phys. Rev. Let. {\bf 69}, 2881
(1992).}

\Title{\vbox{\baselineskip12pt\hbox{UMD-PP 94-132}}}
{\vbox{\centerline{Implications of an Ambiguity in J.S. Bell's}
\vskip9pt
\centerline{ Analysis of the Einstein-Podolsky-Rosen Problem}}}
\vskip9pt

\centerline{Daniel~I.~Fivel~\foot{fivel@umdhep.umd.edu}}

\bigskip\centerline{\it Department of Physics}
\centerline{\it University of Maryland}
\centerline{\it  College Park, MD 20742}

\vskip 13mm

\centerline{\bf Abstract}
\vskip 4mm
An ambiguity is pointed out in J.S. Bell's argument that the
distinction between quantum mechanics and hidden variable theories cannot be
found in the
behavior of single-particle beams.  Within the context of theories for which
states are
unambiguously defined it is shown that the question of whether quantum
mechanics or a locally
realistic theory is valid may indeed be answered by single-particle beam
measurements. It is argued
that  two-particle correlation experiments are required to answer the more
fundamental
question of whether or not the notion of a state can be unambiguously defined.
As a byproduct of the
discussion the general form of completely entangled states is deduced.

\vskip 4mm

\Date{\hfill 4/94}
\vfill\eject

J.S. Bell's 1964 paper~\rBELL~, in which his celebrated inequality first
makes its appearance, proves the inconsistency between quantum mechanics and
locally realistic
theories, i.e. hidden variable theories obeying the Einstein-Podolsky-Rosen
(EPR) postulates~\rEPR~. In a preliminary discussion Bell makes the following
assertion:
``Firstly, there is no difficulty in giving a hidden variable account of spin
measurements on a
single particle." To prove this he considers a particle ``in a pure spin state
labeled by a unit
vector" and proceeds to construct a hidden variable model that reproduces the
quantum mechanical
prediction for the expectation value of the spin in any direction. Bell uses
this construction to
motivate employing {\it two-particle} systems to discover the contradiction
between locally realistic
theories and quantum mechanics  by means of his inequality.

There is, however, an ambiguity in the construction. To see this let us ask
 what one means by saying that a particle is in a pure
state labeled  x. The  operational criterion  is
that we be able to predict with certainty that the particle will pass an
x-filter
(e.g. a Stern-Gerlach filter with magnet orientation indicated by x). There are
 (at least) two methods for doing this: (i) We can first pass the particles
through an $x$-filter.
This procedure is ``invasive" in that we must interact with the particle to
prepare the state.
  (ii) We can exploit correlations. Thus suppose we have a two-particle
system such that there is equal likelihood for a particle to pass any filter x,
but that one can
predict with certainty that a particle will pass $x$ if its partner passes a
filter $x^u$. Here $x^u$ is related to $x$ by some definite rule, and we refer
to the system as a
$u$-{\it correlated} system. Given such a system one prepares a beam in the
state $x$ by admitting
only the partners of particles that have passed an $x^u$-filter. Method (ii)
has the advantage of
being ``non-invasive" in that if measurements are space-like separated events
and we assume Einstein
locality, we can be sure that the state preparation does not disturb the
particle.

Since the method of state preparation is not unique, the question must first be
answered as to
whether the outcome of experiments is sensitive to the method. This is a more
fundamental question
than what the {\it specific} outcome may be since it must be answered before we
use the term
``state". We are going to prove the following two theorems below:

{\it Theorem 1: Quantum mechanics makes the same predictions for both methods
of state
preparation.}

{\it Theorem 2: If a locally realistic theory makes the same predictions for
both methods of state
preparation it will disagree with the quantum mechanical prediction.}

We may draw the following inference from these theorems: For brevity let us say
that dynamics is
``Markovian" if predictions are unambiguous, i.e.\ independent of the method of
state preparation.
(We choose this term because Markov processes are those in which the future is
independent of the
past for a known present.) Classical mechanics has this property, and Theorem 1
shows that
conventional quantum mechanics has it also. Thus a non-Markovian world is going
to be even more
counter-intuitive than quantum mechanics. But Theorem 2 now shows that one
cannot replace quantum
mechanics by a locally realistic theory {\it that is also Markovian} unless we
can prove that the
quantum mechanical predictions for {\it single-particle beams} (e.g.\ Malus'
Law for photons) fails.
In other words {\it within the Markovian framework it is possible to rule out
locally realistic
theories without resorting to sophisticated two-particle correlation
experiments}. Since Bell's
single-particle hidden variable model reproduces the quantum mechanical result
it follows that
this model describes a non-Markovian world. We therefore concur with the
conclusion drawn by
R.Griffiths~\rGRIF~ from his consistent history analysis that the
counter-intuitive
 aspect of quantum mechanics is already revealed in the behavior of single
particle
beams. Thus the two-particle correlation experiments are  seen to be addressed
to a more profound
question, one that applies {\it equally} to classical and quantum mechanics,
namely  whether nature
permits the unambiguous definition of a state. Hence we suggest that such
experiments should be
designed to demonstrate that methods (i) and (ii) give the {\it same} result
rather than focussing
on what results the two methods happen to give individually.

We will use the following notation: We indicate by $P_u(x|y)$ the probability
that one of the
particles in a u-{\it correlated} system passes a y-filter if its partner is
known to have passed an
$x^u$-filter.  Let $p(x|y)$ denote the probability that a particle passes a
y-filter if it
itself has {\it previously} passed an $x$-filter. Then the question of
sensitivity to method of
state preparation may be formulated as:
$$
P_u(x|y) \mathop=\limits^? p(x|y).
\eqno(1)
$$

We first deduce that quantum mechanics verifies (1). Consider quantum
mechanical
particles described by vectors in an N-dimensional Hilbert space with a basis
$|i\rangle$, $i =
1,2,\cdots,N$.

{\it Lemma 1:} A necessary and sufficient condition that a two particle state
$|S\rangle$
exhibit u-correlation is that it be of the form:

$$
|S\rangle = N^{-1/2}\sum_{i=1}^{N}{|i,1\rangle|i^u,2\rangle},
\eqno(2)$$
in which $|i^u\rangle$ is related to $|i\rangle$ by an {\it anti-unitary}
transformation.

The proof of  necessity is given in the appendix. The proof of sufficiency
reveals
the role of anti-unitarity and is as follows: First note that $|S\rangle$ is
independent of the choice of basis. For if $|i\rangle =
\sum_j{A_{ij}|j'\rangle}$ where prime
indicates another basis and $A$ is unitary, then by the anti-unitarity of $u$
we have: $|i^u\rangle = \sum_k{A_{ij}^*|j'^u\rangle}$ whence one verifies that
(2) is
unchanged if $i$ is replaced by $i'$. One may note that in the Bohm
state~\rBOHM~~\rCLAU~ (the spin-0
state of two spin-1/2 particles) the map $u$ is the anti-unitary time-reversal
transformation as may
be demonstrated by observing that the correlated magnet orientations operate
with reversed
currents.

Now let $\pi(x,j) \equiv |x,j\rangle\langle x,j|, \;\; j=1,2$. We then have:
$$
P_u(x|y) = \langle S|\pi(x^u,2)\pi(y,1))|S\rangle / \langle
S|\pi(x^u,2)|S\rangle.
\eqno(3)$$
The denominator on the right is $1/N$. Using independence of the choice of
basis in (2) we select a
basis containing $|y\rangle$. The numerator is then verified to be $|\langle
x^u,2|y^u,2\rangle|^2/N
= |\langle x|y \rangle|^2/N$. But $p(x|y) = |\langle x|y\rangle|^2$ and so we
have proved that
$P_u(x|y) = p(x|y)$ which is Theorem 1 $\;\bullet$

We now turn to Theorem 2: We show that if the quantum mechanical prediction for
the
right side of (1) is assumed, and the left side is computed in a locally
realistic theory, then
the equation will be violated.  According to the
familiar EPR argument we must be able to assign values to determinations of
filter passage.
 Thus we are to assume that there is a set $\Lambda$ whose elements $\lambda$
are the values
of the hidden-variable and that there are subsets $\Lambda_j(x)$ for all filter
labels and $j=1,2$
such that particle $j$ passes an $x$ filter if $\lambda \in \Lambda_j(x)$ and
otherwise does not. If
$\mu(\Lambda)$ is the measure then:
 $$
 P_u(x|y) =
\mu(\Lambda_2(x^u)\cap\Lambda_1(y))/\mu(\Lambda_2(x^u)).
 \eqno(4)$$
The perfect correlation between the two particles means that $\Lambda_2(x^u) =
\Lambda_1(x)$ up to a
set of $\mu$-measure zero so that we may write this as:
$$
P_u(x|y) = \mu(\Lambda_1(x)\cap\Lambda_1(y))/\mu(\Lambda_1(x)).
\eqno(5)$$
Since the quantum mechanical prediction for the right side of (1) is symmetric
in the two arguments,
it follows that $P_u(x|y)$ can reproduce it only if it is also symmetric. But
the numerator on the
right of (5) is symmetric and hence the denominator must be independent of $x$.
Hence it can be
absorbed into the the numerator by redefining $\mu$ and we may write:
$$
P_u(x|y) = \mu(\Lambda_1(x)\cap\Lambda_1(y)).
\eqno(6)$$
Here we may see the striking effect of method (ii) state determination in
constraining the form of $P_u(x|y)$, i.e. we cannot simply take the argument on
the right side to be
an {\it arbitrary} set $\Lambda(x,y)$, but must rather take it to be the {\it
intersection} of sets
depending separately on $x$ and $y$. That Bell's construction~\rBELL~referred
to above is
non-Markovian results from the fact that he utilizes a set which is {\it not}
of this form!

Now suppose that (1) is valid. Then:
$$
 \sup_z |P_u(x|z) - P_u(y|z)| = \sup_z |p(x|z) - p(y|z)|
\eqno(7)$$
We will prove that if the quantum mechanical expression is used for the right
and (6) is
used for the left we will have:

{\it Lemma 2:}
$$
\sup_{z}|p(x|z) - p(y|z)| = (1 - p(x|y))^{1/2},
\eqno(8)$$
$$
\sup_z|P_u(x|z) - P_u(y|z)| = 1 - P_u(x|y).
\eqno(9)$$
To prove (8) note that if $\pi(z) \equiv |z \rangle\langle z|$ we have $p(x|y)
= Tr(\pi(x)\pi(y))$,
and
$$
 \sup_z|Tr(\pi(x)\pi(z)) - Tr(\pi(y)\pi(z))| = \sup_z|\langle z|\pi(x)
-\pi(y)|z \rangle|.
\eqno(10)$$
But this is just the largest eigenvalue~\rBARG~ of $\pi(x) - \pi(y)$. Since the
$\pi$'s  are
projectors:
$$
(\pi (x) - \pi (y))^3 = (1 - |\langle x|y \rangle|^2)(\pi(x) - \pi(y))
\eqno(11)$$
and one reads off the largest eigenvalue to obtain the assertion. To prove (9)
note that in $|\mu(\Lambda(x)\cap\Lambda(z)) - \mu(\Lambda(y)\cap\Lambda(z))|$
the contribution
coming from any overlap of $\Lambda(x)$ and $\Lambda(y)$ will cancel. Hence one
can compute the
supremum as if the sets are disjoint. This occurs when either $z = x$ or $z =
y$ and has
the value given by the right side of (9).

The occurence of the square-root factor in (8) but not in (9) shows that (7)
holds if and only if
$p(x|y)$ is restricted to the values zero or unity. This is true classically
but false in quantum
mechanics, and we have proved Theorem 2 $\bullet$

The reader may have noted that the Bell inequality has not appeared in the
above analysis. It is
concealed in the relationship between (8) and (9) in the following way: One
readily checks that the
quantities  on the left of these two equations are metrics~\rBELIN~, and the
Bell inequality is
just the triangle inequality associated with the metric of (9). The presence of
the square
root on the right side of  (8) that is missing from (9) produces the violation
of that inequality
and a myriad of other contradictions as well, e.g.\ absence of a watch-dog
effect. The fact that the
mathematical origin of the incompatibility between quantum mechanics and
locally realistic theories
is to be found in the metric was noted by the author~\rFIVone~ That crucial
square-root also turns
out to provide interesting insight into the structure of quantum mechanics that
will be discussed
elsewhere~\rFIVtwo~.

 It is to be noted that Theorem 1
is of interest in its own right. It provides a general method for the
construction of perfectly
entangled states and has been used by the author~\rFIVthree~ in the study of
the Bennett-Wiesner
communication scheme~\rBEN~.
\vskip .1in

\centerline{ Appendix --- Proof of necessity in Lemma 1 }
\vskip .1in

Let $|S_x,1\rangle \equiv \langle x^u,2|S\rangle,\;|S_x,2\rangle \equiv \langle
x,1|S\rangle$ (which
makes sense notationally because $|S\rangle$ is a two-particle state). Then
$\langle x,1|S_x,1
\rangle = \langle x^u,2|S_x,2\rangle$. One checks that the perfect correlation
condition can be
expressed in the form:
 $$
\langle S_x,1|x,1\rangle \langle x,1|S_x,1\rangle  =  \langle
S_x,1|S_x,1\rangle  \neq 0,
\eqno(12)
$$
and is equal to the same expression with $1$ replaced by $2.$ Hence one
deduces:
$$
|S_x,1\rangle = \gamma(x)|x,1\rangle  \;\; and \;\;  |S_x,2\rangle  =
\gamma(x)|x^u,2\rangle,
\eqno(13)$$
where $\gamma(x)$ is a non-vanishing complex number.
Then
$$
\gamma(x)\langle y,1|x,1\rangle = \langle y,1|S_x,1\rangle = \langle
x^u,2|<y,1|S\rangle =
\gamma(y)\langle x^u,2|y^u,2\rangle.
 \eqno(14)$$
Multiplying the left and right members by the complex conjugate $\gamma(x)^*$
and noting that
$$
\gamma(x)^*\langle x^u,2|y^u,2\rangle= (\gamma(x)\langle y^u,2|x^u,2\rangle ^*
=
\gamma(y)^*\langle y,1|x,1\rangle,
\eqno(15)$$
it follows that
$$
|\gamma(x)| = |\gamma(y)| \;\; if \;\; \langle x|y \rangle \neq 0,
\eqno(16)$$
so that  $\gamma(x)$ is unimodular up to a constant factor. Hence, by
redefining $|x^u,2\rangle$ to
absorb the unimodular factor $\gamma(x),$ it follows from (14) that the map $u$
is anti-unitary.
One may then select any basis to express $|S\rangle$ in the form
$$
|S\rangle = \sum\limits_{i,j=1}^N{\alpha_{ij}|i,1\rangle |j^u,2\rangle},
\eqno(17)$$
and use $(12)$  to show that $\alpha_{ij} = \delta_{ij}/\sqrt{N}$ thereby
giving $(2).$
$\bullet$

I would like to thank  T.Jacobson, A. Shimony, and C.H. Woo for very
helpful discussions.
\vfill
\listrefs

\end